\RequirePackage{fix-cm}
\documentclass[10pt,a4paper,english,aps,prd,twocolumn,10pt,a4paper,superscriptaddress,nofootinbib,nobibnotes,altaffillsymbol,showpacs,showkeys,floatfix]{revtex4-1}
\usepackage[T1]{fontenc}
\usepackage[utf8]{inputenc}
\usepackage{color}
\usepackage{babel}
\usepackage{mathrsfs}
\usepackage{amsmath}
\usepackage{amssymb}
\usepackage{wasysym}
\usepackage{graphicx}
\usepackage{esint}
\usepackage[unicode=true,
 bookmarks=true,bookmarksnumbered=false,bookmarksopen=true,bookmarksopenlevel=2,
 breaklinks=true,pdfborder={0 0 0},backref=false,colorlinks=true]
 {hyperref}
\hypersetup{pdftitle={The Thirring Model at Finite Density in 2+1 Dimensions with Stochastic Quantization},
 pdfauthor={Jan Pawlowski, Christian Zielinski},
 linkcolor=BlueViolet,anchorcolor=BlueViolet,citecolor=BlueViolet,filecolor=BlueViolet,urlcolor=BlueViolet}

\makeatletter

\pdfpageheight\paperheight
\pdfpagewidth\paperwidth

\newcommand{\lyxdot}{.}

 \@ifundefined{textcolor}{}
 {%
   \definecolor{BLACK}{gray}{0}
   \definecolor{WHITE}{gray}{1}
   \definecolor{RED}{rgb}{1,0,0}
   \definecolor{GREEN}{rgb}{0,1,0}
   \definecolor{BLUE}{rgb}{0,0,1}
   \definecolor{CYAN}{cmyk}{1,0,0,0}
   \definecolor{MAGENTA}{cmyk}{0,1,0,0}
   \definecolor{YELLOW}{cmyk}{0,0,1,0}
 }

\usepackage[usenames,dvipsnames]{xcolor}

\allowdisplaybreaks
\usepackage{slashed}
\usepackage{bbm}

\@ifundefined{showcaptionsetup}{}{%
 \PassOptionsToPackage{caption=false}{subfig}}
\usepackage{subfig}
\makeatother

\begin{document}

\title{Thirring model at finite density in $2+1$ dimensions\\
 with stochastic quantization}

\author{Jan M.~Pawlowski}

\affiliation{Institut für Theoretische Physik, Universität Heidelberg, Philosophenweg
16, 69120 Heidelberg, Germany}

\affiliation{ExtreMe Matter Institute EMMI, GSI, Planckstraße 1, D-64291 Darmstadt,
Germany}

\author{Christian Zielinski}

\altaffiliation{Permanent address: Division of Mathematical Sciences, Nanyang Technological University, Singapore 637371}

\affiliation{Institut für Theoretische Physik, Universität Heidelberg, Philosophenweg
16, 69120 Heidelberg, Germany}

\date{\today}
\begin{abstract}
We consider a generalization of the Thirring model in $2+1$ dimensions
at finite density. We employ stochastic quantization and check for
the applicability in the finite density case to circumvent the sign
problem. To this end we derive analytical results in the heavy dense
limit and compare with numerical ones obtained from a complex Langevin
evolution. Furthermore we make use of indirect indicators to check
for incorrect convergence of the underlying complex Langevin evolution.
The method allows the numerical evaluation of observables at arbitrary
values of the chemical potential. We evaluate the results and compare
to the $\left(0+1\right)$--dimensional case. 
\end{abstract}



\maketitle

\section{Introduction \label{sec:Introduction}}

The sign problem remains one of the biggest challenges of lattice
field theory until this day. It is caused by a highly oscillatory
and complex path integral measure after introducing a chemical potential
$\mu$. This renders many theories, like four-dimensional quantum
chromodynamics (QCD), inaccessible by common simulation algorithms
in wide regions of the phase diagram. Many solutions have been proposed,
see e.g.,~\cite{deForcrand:2010ys,Lombardo:2005gj,Aarts:2013bla,Aarts:2009yj,Delgado:2012tm,Schmidt:2012uy,Alexandru:2010yb,Alexandru:2005ix}.
However, reliable numerical calculations in theories with a severe
sign problem remain extremely challenging.

One of the approaches to the sign problem is stochastic quantization
\cite{Parisi:1980ys}, i.e., a complex Langevin evolution \cite{Parisi:1984cs}.
For a review of stochastic quantization see e.g., \cite{Damgaard:1987rr},
and for its application to the finite density case refer to \cite{Aarts:2013bla}.
It has been applied to many theories suffering from a severe sign
problem, often successfully \cite{Karsch:1985cb,Aarts:2011zn,Bilic:1987fn,Aarts:2008rr,Aarts:2010gr,Aarts:2008wh,Aarts:2009hn}.
It was also applied outside the context of finite density calculations,
e.g., quantum fields in nonequilibirum \cite{Berges:2005yt} and in
real time \cite{Berges:2006xc,Berges:2007nr}. However, there are
cases known where the complex Langevin evolution converges towards
unphysical fixed points \cite{Ambjorn:1986fz,Aarts:2010aq}. Starting
from early studies of complex Langevin evolutions \cite{Hamber:1985qh,Flower:1986hv,Ilgenfritz:1986cd}
until this day, the convergence properties of Langevin equations generalized
to the complex case are not well understood. In this work we focus
on the stochastic quantization of the Thirring model at finite density
in $2+1$ dimensions, cf.~\cite{Spielmann10}. This serves as a toy
model for the matter sector of QCD. Furthermore, the $\left(2+1\right)$--dimensional
model appears in effective theories of high temperature superconductors
and graphene, see e.g., references given in \cite{Gies:2010st}.

Here we extend the $\left(0+1\right)$--dimensional studies carried
out in \cite{Pawlowski:2013pje}. In $2+1$ dimensions we lose the
analytic benchmarks which facilitated the interpretation of the numerical
results with stochastic quantization in $0+1$ dimensions. Still,
we can exploit the similarities to the lower-dimensional case. Further
benchmarks of our numerical results are given by those in the heavy
dense limit, as introduced in \cite{Bender:1992gn}. This limit describes
the regime of large fermion masses and large chemical potentials.
In addition, we evaluate indirect indicators, namely the consistency
requirements presented in \cite{Aarts:2009uq,Aarts:2011ax,Aarts:2011sf}
and the analyticity of the fermion condensate at $\mu^{2}=0$, cf.~\cite{Aarts:2011zn}.

The paper is organized as follows: In Sec.~\ref{sec:The-Thirring-Model}
we introduce a generalized Thirring model in the continuum and on
the lattice. We also formulate the associated Langevin equation. In
Sec.~\ref{sec:Analytical-Results} we employ the heavy dense limit
to derive numerous analytical results. At the end of the section we
introduce indirect indicators of correct convergence, namely analyticity
of observables at $\mu^{2}=0$ and a set of consistency conditions.
In Sec.~\ref{sec:Numerical-Results} we discuss numerical results
and use the analytical results among other indicators as a benchmark
to evaluate the complex Langevin evolution. Finally in Sec.~\ref{sec:Conclusions}
we discuss and summarize our findings.

\global\long\def\Tr{\operatorname{Tr}}

\global\long\def\myRe{\operatorname{Re}}

\global\long\def\myIm{\operatorname{Im}}

\global\long\def\matrixOne{\mathbbm{1}}

\global\long\def\ii{\textrm{i}}

\section{The Generalized Thirring Model \label{sec:The-Thirring-Model}}

\subsection{Continuum formulation \label{sub:Continuum-formulation}}

We begin with a short recapitulation of the model, which we introduced
in \cite{Pawlowski:2013pje}. It is a generalization of the historical
$1+1$ dimensional Thirring model \cite{Thirring:1958in}, but formulated
in $d$ dimensions with $N_{f}$ fermion flavors at finite density.
The Euclidean Lagrangian reads
\begin{multline}
\mathscr{L}_{\Psi}=\sum_{i=1}^{N_{f}}\overline{\Psi}_{i}\left(\slashed{\partial}+m_{i}+\mu_{i}\gamma_{0}\right)\Psi_{i}\\
+\frac{g^{2}}{2N_{f}}\left(\sum_{i=1}^{N_{f}}\overline{\Psi}_{i}\gamma_{\nu}\Psi_{i}\right)^{2}.\label{eq:TMLagrangeHist}
\end{multline}
Here $i=1,\ldots,N_{f}$ is a flavor index, $m_{i}$ and $\mu_{i}$
are the bare mass and bare chemical potential of the respective flavor
and $g^{2}$ is the bare coupling strength. The $\gamma$ matrices
satisfy $\left\{ \gamma_{\mu},\gamma_{\nu}\right\} =2\delta_{\mu\nu}\matrixOne$.

In $d=2+1$ dimensions, $\overline{\Psi}$ and $\Psi$ denote four-component
spinors. While it is possible to use an irreducible two-dimensional
representation of the Dirac algebra, this representation does not
allow for a chiral symmetry in the massless case. See \cite{Gies:2010st}
for details and a discussion of symmetries. It is noted that the model
also shows breaking of chiral symmetry at vanishing chemical potential
\cite{Christofi:2007ye}.

We reformulate the generalized Thirring model using an auxillary field
and integrate out the fermionic degrees of freedom. We find for the
partition function
\begin{align}
Z & =\intop\mathscr{D}A\,\left(\prod_{i}\det K_{i}\right)\, e^{-S_{A}}=\intop\mathscr{D}A\, e^{-S_{\textrm{eff}}},\nonumber \\
S_{A} & =N_{f}\beta\intop_{0}^{1/T}\textrm{d}t\intop\textrm{d}^{d-1}\mathbf{x}\, A_{\nu}^{2}\label{eq:ContPartFunc}
\end{align}
with temperature $T$, fermionic term $K_{i}=\slashed{\partial}+\ii\slashed{A}+m_{i}+\mu_{i}\gamma_{0}$
and $S_{\textrm{eff}}=S_{A}-\sum_{i}\Tr\log K_{i}$. The fermion determinant
obeys
\begin{equation}
\det K_{i}\left(\mu\right)=\left[\det K_{i}\left(-\mu^{\star}\right)\right]^{\star},\label{eq:Fermion-det-symmetry}
\end{equation}
yielding in general a complex action. The sign problem can be avoided
by taking the absolute value of the fermion determinant. This corresponds
to an isospin chemical potential and is also referred to as the phase-quenched
case. We also note that observables shall become independent of the
chemical potential $\mu$ up to some threshold $\mu_{c}$ in the zero-temperature
limit. This goes under the name of the Silver Blaze problem \cite{Cohen:2003kd}.

\subsection{Lattice formulation \label{sub:Lattice-formulation}}

We consider the generalized Thirring model in $2+1$ dimensions on
a space-time lattice with $N_{t}$ time slices and $N_{s}$ slices
in spatial direction. Furthermore, we require that $N_{t}$ be even
\cite{Pawlowski:2013pje}. The spatial volume and the space-time volume
are denoted by
\begin{equation}
V=N_{s}^{2},\qquad\Omega=N_{t}N_{s}^{2}.\label{eq:Volume-Def}
\end{equation}
We use staggered fermions \cite{Kogut:1974ag,Banks:1975gq,Banks:1976ia,Susskind:1976jm},
where the number of staggered fermion fields---denoted also as lattice
flavors---is given by $\mathcal{N}$. The introduction of a chemical
potential follows the prescription by Hasenfratz and Karsch \cite{Hasenfratz:1983ba}.
In the following, all dimensionful quantities are measured in appropriate
powers of $a$, so that we only deal with dimensionless parameters.

In three dimensions, $\mathcal{N}$ staggered fermion flavors correspond
to $N_{f}=2\mathcal{N}$ continuum flavors \cite{Spielmann10} as
each staggered fermion field encodes two tastes. In principle the
partition function for a single continuum flavor can be obtained by
taking the square root of the fermion determinant, but it is still
under debate if this rooting prescription is consistent \cite{Sharpe:2006re}.

In $d=2+1$ dimensions (where generalizations to arbitrary $d$ are
evident), the lattice action we employ reads
\begin{equation}
S=\sum_{x,y,i}\overline{\chi}_{i}\left(x\right)K_{i}\left(x,y\right)\chi_{i}\left(y\right)+\frac{\mathcal{N}\beta}{2}\sum_{x,\nu}A_{\nu}^{2}\left(x\right).
\end{equation}
Here $\overline{\chi}_{i}$ and $\chi_{i}$ denote staggered fermion
fields with flavor index $i=1,\ldots,\mathcal{N}$, and the sums extend
over $x,y=1,\ldots,\Omega$ and $\nu=0,\ldots,d-1$. The fermion matrix
reads
\begin{multline}
K_{i}\left(x,y\right)=\frac{1}{2}\sum_{\nu=0}^{d-1}\varepsilon_{\nu}\left(x\right)\left[\left(1+\ii A_{\nu}\left(x\right)\right)e^{\mu_{i}\delta_{\nu0}}\delta_{x+\hat{\nu},y}\right.\\
\left.-\left(1-\ii A_{\nu}\left(y\right)\right)e^{-\mu_{i}\delta_{\nu0}}\delta_{x-\hat{\nu},y}\right]+m_{i}\delta_{xy}\label{eq:Fermion-Matrix}
\end{multline}
with staggered phase factor 
\begin{equation}
\varepsilon_{\nu}(x)=\left(-1\right)^{\sum_{i=0}^{\nu-1}x_{i}}
\end{equation}
and $\hat{\nu}$ denoting a unit vector in $\nu$ direction, cf.~\cite{DelDebbio:1995zc,Spielmann10}.
We impose periodic boundary conditions in spatial and antiperiodic
boundary conditions in temporal direction. The lattice partition function
is, like in the continuum in \eqref{eq:ContPartFunc}, given by
\begin{equation}
Z=\intop_{-\infty}^{\infty}\prod_{x,v}\textrm{d}A_{v}\left(x\right)\,\left(\prod_{i}\det K_{i}\right)\, e^{-S_{A}},\label{eq:Lattice-Part-Func}
\end{equation}
where $S_{A}=\frac{1}{2}\mathcal{N}\beta\sum_{x,\nu}A_{\nu}^{2}\left(x\right)$.
The central observables in our analysis are the fermion density, the
fermion condensate, the energy density and the phase factor of the
fermion determinant. In the following, sums over flavor indices are
not implied. The fermion density of flavor $i$ is given by
\begin{equation}
\left\langle n_{i}\right\rangle =\frac{1}{\Omega}\left(\frac{\partial\log Z}{\partial\mu_{i}}\right)_{V,T}=\frac{1}{\Omega}\left\langle \Tr\left(\frac{\partial K_{i}}{\partial\mu_{i}}K_{i}^{-1}\right)\right\rangle .\label{eq:Formula-density}
\end{equation}
The fermion condensate follows from
\begin{equation}
\left\langle \overline{\chi}_{i}\chi_{i}\right\rangle =\frac{1}{\Omega}\left(\frac{\partial\log Z}{\partial m_{i}}\right)_{V,T,\mu_{i}}=\frac{1}{\Omega}\left\langle \Tr K_{i}^{-1}\right\rangle ,\label{eq:Formula-condensate}
\end{equation}
and the energy density reads
\begin{equation}
\left\langle \varepsilon_{i}\right\rangle =-\left(\frac{\partial\log Z}{\partial N_{t}}\right)_{V,\mu_{i}}+\mu_{i}\left\langle n_{i}\right\rangle .\label{eq:Formula-energy}
\end{equation}
Usually we normalize the latter one to $\left\langle \varepsilon_{i}\right\rangle \left(\mu=0\right)=0$.

The phase factor of the fermion determinant is defined by $\exp\left(\ii\phi\right)=\det K/\left|\det K\right|$
with phase $\phi$. For $\mathcal{N}$ degenerated staggered fermion
flavors we set $K_{i}=K$. The expectation value of $\exp\left(\ii\mathcal{N}\phi\right)$
\cite{Han:2008xj,Andersen:2009zm} follows in the $\mathcal{N}$ flavor
phase-quenched theory from
\begin{equation}
\left\langle e^{\ii\mathcal{N}\phi}\right\rangle _{\mathcal{N}}^{\textrm{pq}}=\frac{Z_{\mathcal{N}}}{Z_{\mathcal{N}}^{\textrm{pq}}}\in\left[0,1\right],
\end{equation}
where $Z_{\mathcal{N}}$ is given by \eqref{eq:Lattice-Part-Func}
and $Z_{\mathcal{N}}^{\textrm{pq}}$ denotes the phase-quenched partition
function. The expectation value of the fermion phase factor is a measure
of the sign problem, where smaller values indicate a more severe problem.

\subsection{Complex Langevin equation \label{sub:Formulating-the-Langevin-Eq}}

To deal with the associated sign problem we apply stochastic quantization
\cite{Parisi:1980ys}, namely a complex Langevin evolution \cite{Parisi:1984cs},
to the Thirring model at finite density. To this end we determine
the stationary solution of the corresponding Langevin equation, which
reads
\begin{equation}
\frac{\partial}{\partial\Theta}A_{\nu}\left(x,\Theta\right)=-\frac{\delta S_{\textrm{eff}}\left[A\right]}{\delta A_{\nu}\left(x,\Theta\right)}+\sqrt{2}\,\eta_{\nu}\left(x,\Theta\right).\label{eq:Cont-Langevin-eq}
\end{equation}
Here $\Theta$ denotes fictitious time, and $\eta_{\nu}\left(x,\Theta\right)$
is a Gaussian noise with
\begin{align}
\left\langle \eta_{\nu}\left(x,\Theta\right)\right\rangle  & =0,\nonumber \\
\left\langle \eta_{\nu}\left(x,\Theta\right)\eta_{\sigma}\left(x',\Theta'\right)\right\rangle  & =\delta_{\nu\sigma}\delta\left(x-x'\right)\delta\left(\Theta-\Theta'\right).
\end{align}
We use an adaptive first-order stepsize integration scheme \cite{Ambjorn:1985iw,Aarts:2009dg,Spielmann10},
where the stepsize $\epsilon_{L}$ is adjusted with respect to the
modulus of the drift term.

For the numerical treatment of the Langevin equation we use a first-order
integration scheme. Our discretization of \eqref{eq:Cont-Langevin-eq}
reads
\begin{multline}
A_{\nu}\left(x,\Theta+\epsilon_{L}\right)=\\
A_{\nu}\left(x,\Theta\right)+\epsilon_{L}D_{\nu}\left(x,\Theta\right)+\sqrt{2\epsilon_{L}}\,\eta_{\nu}\left(x,\Theta\right),
\end{multline}
where $D_{\nu}\left(x,\Theta\right)=-\textrm{d}S_{\textrm{eff}}/\textrm{d}A_{\nu}\left(x,\Theta\right)$
is the drift term and $\epsilon_{L}$ the (adaptive) integration stepsize.
The drift term takes the explicit form
\begin{multline}
D_{\nu}\left(x,\Theta\right)=-\mathcal{N}\beta A_{\nu}\left(x,\Theta\right)\\
+\frac{\ii}{2}\varepsilon_{\nu}\left(x\right)\sum_{i}K_{i}^{-1}\left(x+\hat{\nu},x\right)e^{\mu_{i}\delta_{v0}}\\
+\frac{\ii}{2}\varepsilon_{\nu}\left(x+\hat{\nu}\right)\sum_{i}K_{i}^{-1}\left(x,x+\hat{\nu}\right)e^{-\mu_{i}\delta_{v0}},
\end{multline}
cf.~\cite{Spielmann10}. The stepsize is updated after each integration
step according to
\begin{equation}
\epsilon_{L}\equiv\epsilon_{L}\left(\Theta\right)=\frac{\delta}{\max_{x,\nu}\left|D_{\nu}\left(x,\Theta\right)\right|}.\label{eq:Adapative-stepsize}
\end{equation}
Typically we use $\delta=10^{-2}$, see also \cite{Pawlowski:2013pje}.

For a correctly converging complex Langevin evolution we can generalize
the real noise to an imaginary noise in \eqref{eq:Cont-Langevin-eq}
while keeping expectation values unchanged \cite{Spielmann10}. We
parametrize the complex noise using the replacement
\begin{equation}
\eta_{\nu}\left(x,\Theta\right)\to\sqrt{\mathcal{I}+1}\myRe\eta_{\nu}\left(x,\Theta\right)+\ii\sqrt{\mathcal{I}}\myIm\eta_{\nu}\left(x,\Theta\right)
\end{equation}
with $\mathcal{I}\geqslant0$. Furthermore, we require that
\begin{multline}
\left\langle \myRe\eta_{\nu}\left(x,\Theta\right)\myRe\eta_{\sigma}\left(x',\Theta'\right)\right\rangle \\
=\left\langle \myIm\eta_{\nu}\left(x,\Theta\right)\myIm\eta_{\sigma}\left(x',\Theta'\right)\right\rangle \\
=\delta_{\nu\sigma}\delta\left(x-x'\right)\delta\left(\Theta-\Theta'\right)
\end{multline}
and
\begin{equation}
\left\langle \myRe\eta_{\nu}\left(x,\Theta\right)\myIm\eta_{\sigma}\left(x',\Theta'\right)\right\rangle =0.
\end{equation}
While maintaining numerical stability, we check if observables turn
out to be independent of $\mathcal{I}$.

\section{Analytical Results \label{sec:Analytical-Results}}

In the following we derive approximate expressions for some observables
in the lattice Thirring model in $d$ dimensions. We begin with a
hopping parameter expansion and then take the so-called heavy dense
limit in Sec.~\ref{sec:Heavy-dense-limit}. This renders the model
effectively one-dimensional and allows us to obtain a simple expression
for the fermion determinant. Later we discuss an extended version
in Sec.~\ref{sec:Symmetric-Heavy-Dense}.

\subsection{Hopping parameter expansion}

Applying a hopping parameter expansion to the fermion determinant
in \eqref{eq:Lattice-Part-Func} yields
\begin{equation}
\frac{\det K}{m^{\Omega}}=\prod_{\ell}\prod_{\{C_{\ell}\}}\left(1-\kappa^{\ell}\gamma_{C_{\ell}}\mathbf{P}_{C_{\ell}}\right).
\end{equation}
Here $\kappa=1/\left(2m\right)$ is the hopping parameter, $\left\{ C_{\ell}\right\} $
are closed contours of perimeter $\ell$, $n$ is the number of times
a given contour is traced out and $\mathbf{P}_{C_{\ell}}$ is the
product of hopping terms $M\left(x,y\right)$ at $\mu=0$ along the
given contour $C_{\ell}$. The matrix elements $M\left(x,y\right)$
read
\begin{multline}
M\left(x,y\right)=-\sum_{\nu=0}^{d-1}\varepsilon_{\nu}\left(x\right)\left[\left(1+\ii A_{\nu}\left(x\right)\right)\delta_{x+\hat{\nu},y}\right.\\
\left.-\left(1-\ii A_{\nu}\left(y\right)\right)\delta_{x-\hat{\nu},y}\right],
\end{multline}
compare with the fermion matrix in \eqref{eq:Fermion-Matrix}. Furthermore,
we represent the dependence on the chemical potential explicitly via
\begin{equation}
\gamma_{C_{\ell}}=\left[-\exp\left(\mu N_{t}\right)\right]^{\rightturn C_{\ell}},\label{eq:Gamma-Factor}
\end{equation}
where
\begin{equation}
\rightturn C_{\ell}\in\mathbb{Z}\label{eq:Winding-Number}
\end{equation}
denotes the temporal winding number of the path $C_{\ell}$ counted
in positive direction. The minus sign in \eqref{eq:Gamma-Factor}
stems from antiperiodic temporal boundary conditions.

\subsection{Heavy dense limit \label{sec:Heavy-dense-limit}}

The heavy dense limit projects out the leading contributions of the
hopping parameter expansion in the limit of a large mass $m$ and
large chemical potential $\mu$. This limit was introduced in \cite{Bender:1992gn}
and employed e.g., in \cite{Blum:1995cb,Engels:1999tz,Aarts:2001dz,Fukushima:2006uv,Hofmann:2003vv,DePietri:2007ak,Aarts:2008rr}.
It is defined by
\begin{equation}
\kappa\to0,\quad\mu\to\infty,\quad\kappa e^{\mu}\;\textrm{fixed}.\label{eq:Def-HD-limit}
\end{equation}
In this regime the fermion determinant is dominated by contributions
from Polyakov loops in positive time direction with $\rightturn C=1$,
see \eqref{eq:Winding-Number}. Due to the absence of spatial paths,
the model is effectively one dimensional. The fermionic contribution
$\det\mathcal{K}$ in this limit reads
\begin{equation}
\frac{\det\mathcal{K}}{m^{\Omega}}=\prod_{C\in\mathbb{P}}\left(1+\xi\mathbf{P}_{C}\right),
\end{equation}
where $\mathbb{P}$ denotes the set of Polyakov loops and
\begin{equation}
\xi\equiv\zeta^{N_{t}},\qquad\zeta\equiv\kappa e^{\mu}.
\end{equation}
Let $\mathscr{P}_{\mathbf{x}}$ denote a Polyakov loop
\begin{equation}
\mathscr{P}_{\mathbf{x}}=\prod_{t=1}^{N_{t}}\left(1+\ii A_{0}\left(t,\mathbf{x}\right)\right)
\end{equation}
starting and ending in the space point $\mathbf{x}$. We can then
express the fermion determinant in the regime of $m$ and $\mu$ being
large by
\begin{equation}
\det\mathcal{K}=m^{\Omega}\prod_{\mathbf{x}}\left(1+\xi\mathscr{P}_{\mathbf{x}}\right),\label{eq:Heavy-Dense-det}
\end{equation}
where $\Omega$ is the space-time volume defined in \eqref{eq:Volume-Def}.
Note that in this limit the relation in \eqref{eq:Fermion-det-symmetry}
is violated. Due to this approximation we will find $\left\langle n\right\rangle _{\mu=0}\to0$
only for $m\to\infty$. In Sec.~\ref{sec:Symmetric-Heavy-Dense}
we introduce a modified version of this limit, which preserves this
relation.

\subsection{The case of one flavor \label{sub:Single-flavor-partition-function}}

\begin{figure}
\subfloat[Plot of the fermion density $\left\langle n\right\rangle $.]{\includegraphics[width=1\columnwidth]{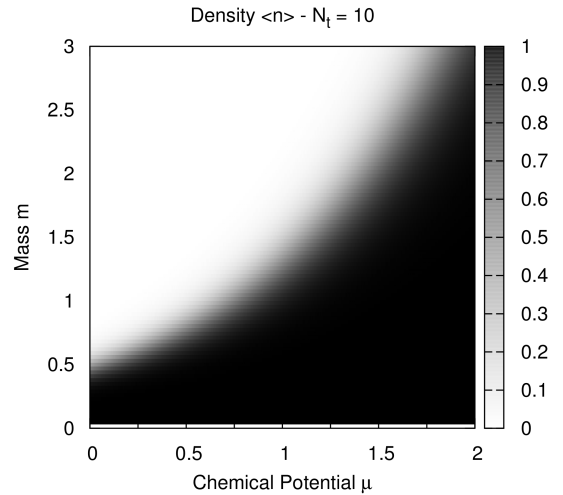}

}

\subfloat[Plot of the fermion condensate $\left\langle \overline{\chi}\chi\right\rangle $.]{\includegraphics[width=1\columnwidth]{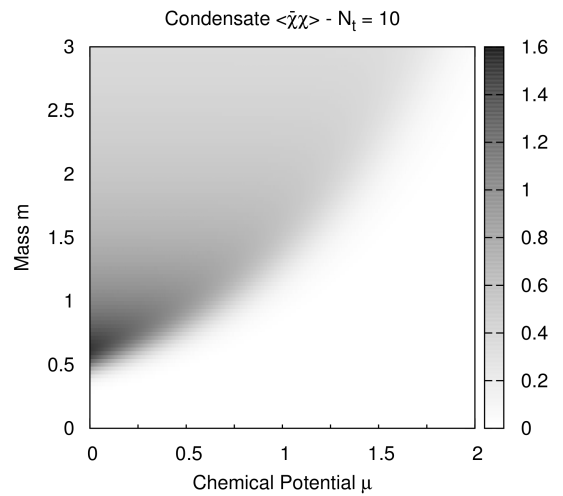}

}

\caption{\label{fig:Phase-diagram} The phase structure in the heavy dense
limit for one flavor.}
\end{figure}

By replacing the full fermion determinants in \eqref{eq:Lattice-Part-Func}
with the simpler expression in \eqref{eq:Heavy-Dense-det}, we can
derive analytical expressions for several observables of interest.
We begin with one flavor and later generalize to more flavors. In
the heavy dense limit the partition function can be integrated exactly,
and we find
\begin{equation}
\mathcal{Z}_{1}=\intop_{-\infty}^{\infty}\prod_{x,v}\textrm{d}A_{v}\left(x\right)\,\det\mathcal{K}\, e^{-S_{A}}=\frac{\left(1+\xi\right)^{V}}{\left(2\kappa\right)^{\Omega}}\left(\frac{2\pi}{\beta}\right)^{\frac{\Omega d}{2}}.\label{eq:Heavy-Dense-Part-Func}
\end{equation}
Using \eqref{eq:Formula-density} and \eqref{eq:Formula-condensate}
the fermion density and condensate read
\begin{equation}
\left\langle n\right\rangle =\frac{1}{1+\frac{1}{\xi}},\qquad\left\langle \overline{\chi}\chi\right\rangle =\frac{2\kappa}{1+\xi}.
\end{equation}
We point out that the exact $\left(0+1\right)$--dimensional results
derived in \cite{Pawlowski:2013pje} reproduce the above results in
the corresponding limit. Applying \eqref{eq:Formula-energy}, the
normalized energy density reads
\begin{equation}
\left\langle \varepsilon\right\rangle =\frac{\left(\xi-\kappa^{1/T}\right)\log\kappa^{-1}}{\left(1+\kappa^{1/T}\right)\left(1+\xi\right)},
\end{equation}
where we identify the temperature with $T=N_{t}^{-1}$. Furthermore,
we find
\begin{equation}
c_{V}=\frac{1}{VT^{2}}\left(\frac{\partial^{2}\log\mathcal{Z}_{1}}{\partial N_{t}^{2}}\right)_{V,\mu}=\frac{\xi\log^{2}\zeta}{T^{2}\left(1+\xi\right)^{2}}>0
\end{equation}
for the heat capacity. The corresponding mechanical equation of state
$PV=T\log\mathcal{Z}_{1}$ takes the form
\begin{equation}
P=\log\left[\frac{\left(1+\xi\right)^{1/T}}{2\kappa}\left(\frac{2\pi}{\beta}\right)^{d/2}\right]
\end{equation}
with $P$ denoting pressure.

Figure \ref{fig:Phase-diagram} shows the phase structure in the heavy
dense limit. Note that $\beta$ dropped out in most considered observables.
For large $N_{t}$ we find two well-separated phases, where the system
is in a condensed phase for large $\mu$.

In the limit of vanishing temperature we find
\begin{align}
\left\langle n\right\rangle _{T=0} & =\Theta\left(\mu-\mu_{c}\right),\nonumber \\
\left\langle \overline{\chi}\chi\right\rangle _{T=0} & =2\kappa\,\Theta\left(\mu_{c}-\mu\right),\\
\left\langle \varepsilon\right\rangle _{T=0} & =\mu_{c}\,\Theta\left(\mu-\mu_{c}\right),\nonumber 
\end{align}
where $\Theta$ is the Heaviside step function and the critical chemical
potential onset is found to be $\mu_{c}=\log\left(2m\right)$. Note
that $\mu_{c}>0$ in the heavy dense regime. The model clearly exhibits
Silver Blaze behavior as mentioned in Sec.~\ref{sec:Introduction}.

\subsection{The case of two flavors \label{sub:Partition-function-two-flavors}}

We continue with determining the partition function for two flavors
in the heavy dense limit. For nondegenerated flavors we label the
parameters $\kappa$, $\mu$ and $\xi$ with a flavor index $i\in\left\{ 1,2\right\} $
and find after an exact integration 
\begin{align}
\mathcal{Z}_{2} & =\intop_{-\infty}^{\infty}\prod_{x,v}\textrm{d}A_{v}\left(x\right)\,\det\mathcal{K}_{1}\,\det\mathcal{K}_{2}\, e^{-S_{A}}\nonumber \\
 & =\frac{\left(1+\xi_{1}+\xi_{2}+\xi_{1}\xi_{2}\Delta\right)^{V}}{\left(4\kappa_{1}\kappa_{2}\right)^{\Omega}}\left(\frac{\pi}{\beta}\right)^{\frac{\Omega d}{2}},
\end{align}
where we introduce
\begin{equation}
\Delta=\left(1-\frac{1}{2\beta}\right)^{N_{t}}
\end{equation}
for brevity. For simplicity we quote the observables only in the case
of degenerated flavors and a common chemical potential, i.e., $\kappa_{i}=\kappa$,
$\mu_{i}=\mu$ and $\xi_{i}=\xi$. The generalization to nondegenerated
flavors is straightforward. The total fermion density is given by
\begin{equation}
\left\langle n\right\rangle =\frac{2\xi\left(1+\xi\Delta\right)}{1+2\xi+\xi^{2}\Delta}
\end{equation}
with $\left\langle n\right\rangle \to2$ for $\mu\to\infty$ (if $\beta\neq1/2$),
while the fermion condensate reads
\begin{equation}
\left\langle \overline{\chi}\chi\right\rangle =\frac{4\kappa\left(1+\xi\right)}{1+2\xi+\xi^{2}\Delta}.
\end{equation}
The mechanical equation of state takes the form
\begin{equation}
P=\log\left[\frac{\left(1+2\xi+\xi^{2}\Delta\right)^{1/T}}{4\kappa^{2}}\left(\frac{\pi}{\beta}\right)^{d/2}\right].
\end{equation}
Like in $0+1$ dimensions \cite{Pawlowski:2013pje} we find a plateau
in the density, condensate and energy density for $\mathcal{N}=2$.
They become visible for couplings of the order $\beta\approx1/2$,
see Fig.~\ref{fig:Multiflavor}. For the special case of $\beta=1/2$
(i.e.~$\Delta=0$), the density never goes into full saturation for
$\mu\to\infty$, but is stuck on the plateau. These structures can
be understood in the $\left(0+1\right)$--dimensional continuum case
\cite{Pawlowski:2013pje}, see also \cite{Banerjee:2010kc}.

\paragraph*{Phase-quenched case.}

We also consider the partition function with a phase-quenched fermion
determinant as mentioned in Sec.~\ref{sub:Lattice-formulation}.
Then in the case of two degenerated flavors, the partition function
reads
\begin{align}
\mathcal{Z}_{2}^{\textrm{pq}} & =\intop_{-\infty}^{\infty}\prod_{x,v}\textrm{d}A_{v}\left(x\right)\,\left|\det\mathcal{K}\right|^{2}\, e^{-S_{A}}\nonumber \\
 & =\frac{1}{\left(2\kappa\right)^{2\Omega}}\left[\left(1+2\xi+\xi^{2}\Delta_{\textrm{pq}}\right)\left(\frac{\pi}{\beta}\right)^{N_{t}d/2}\right]^{V}
\end{align}
with
\begin{equation}
\Delta_{\textrm{pq}}=\left(1+\frac{1}{2\beta}\right)^{N_{t}}.
\end{equation}
The observables are of the same form as in the full theory with $\Delta$
replaced by $\Delta_{\textrm{pq}}$.

\paragraph*{Phase factor.}

The previous results allow us to derive an analytical expression for
the expectation value of the phase factor of the fermion determinant
in the two-flavor theory. It serves as a measure for the severity
of the sign problem. We find
\begin{equation}
\left\langle e^{2\ii\phi}\right\rangle _{\mathcal{N}=2}^{\textrm{pq}}=\frac{\mathcal{Z}_{2}}{\mathcal{Z}_{2}^{\textrm{pq}}}=e^{-V/V_{0}},
\end{equation}
where we define
\begin{equation}
V_{0}=\log\left(\frac{1+2\xi+\xi^{2}\Delta_{\textrm{pq}}}{1+2\xi+\xi^{2}\Delta}\right).
\end{equation}
As expected, the sign problem gets more severe for larger lattices.
In the limit of large chemical potentials the expectation value approaches
\begin{equation}
\lim_{\mu\to\infty}\left\langle e^{2\ii\phi}\right\rangle _{\mathcal{N}=2}^{\textrm{pq}}=\left(\frac{2\beta-1}{2\beta+1}\right)^{\Omega}.
\end{equation}
We see that the choice of $\beta$ has a significant impact on the
severity of the sign problem. In the heavy dense limit we find that
it is most severe for $\beta=1/2$.

\subsection{The case of $\mathcal{N}$ flavors \label{sub:Partition-function-many-flavors}}

Finally, we generalize \eqref{eq:Heavy-Dense-Part-Func} to an arbitrary
number of $\mathcal{N}$ degenerated flavors by raising the fermion
determinant to the power of $\mathcal{N}$ in the partition function.
We obtain
\begin{multline}
\mathcal{Z}_{\mathcal{N}}=\frac{1}{\left(2\kappa\right)^{\mathcal{N}\Omega}}\left(\frac{2\pi}{\mathcal{N}\beta}\right)^{\frac{\Omega d}{2}}\\
\times\left[\sum_{k=0}^{\mathcal{N}}\binom{\mathcal{N}}{k}\Lambda_{k}\xi^{k}U^{N_{t}}\left(\frac{1-k}{2},\frac{3}{2},\frac{\mathcal{N}\beta}{2}\right)\right]^{V}
\end{multline}
in the heavy dense limit, where we introduce
\begin{equation}
\Lambda_{k}=\left(\frac{2}{\mathcal{N}\beta}\right)^{\frac{N_{t}\left(k-1\right)}{2}}.
\end{equation}
Here $\Gamma\left(z\right)$ denotes Euler's gamma function, and $U\left(a,b,z\right)$
is the confluent hypergeometric function of the second kind \cite{abramowitz1964handbook},
also known as Kummer's function. We do not want to quote the observables
here explicitly, but as required they reproduce previous results for
$\mathcal{N}=1,2$. From the expressions for the density and condensate
we obtain the relation
\begin{eqnarray}
\frac{\left\langle \overline{\chi}\chi\right\rangle }{\mathcal{N}} & = & 2\kappa\left(1-\frac{\left\langle n\right\rangle }{\mathcal{N}}\right).
\end{eqnarray}
In this general case we find up to $\mathcal{N}-1$ intermediate plateaus
in the observables.

\subsection{Symmetric heavy dense limit \label{sec:Symmetric-Heavy-Dense}}

\begin{figure}
\includegraphics[width=1\columnwidth]{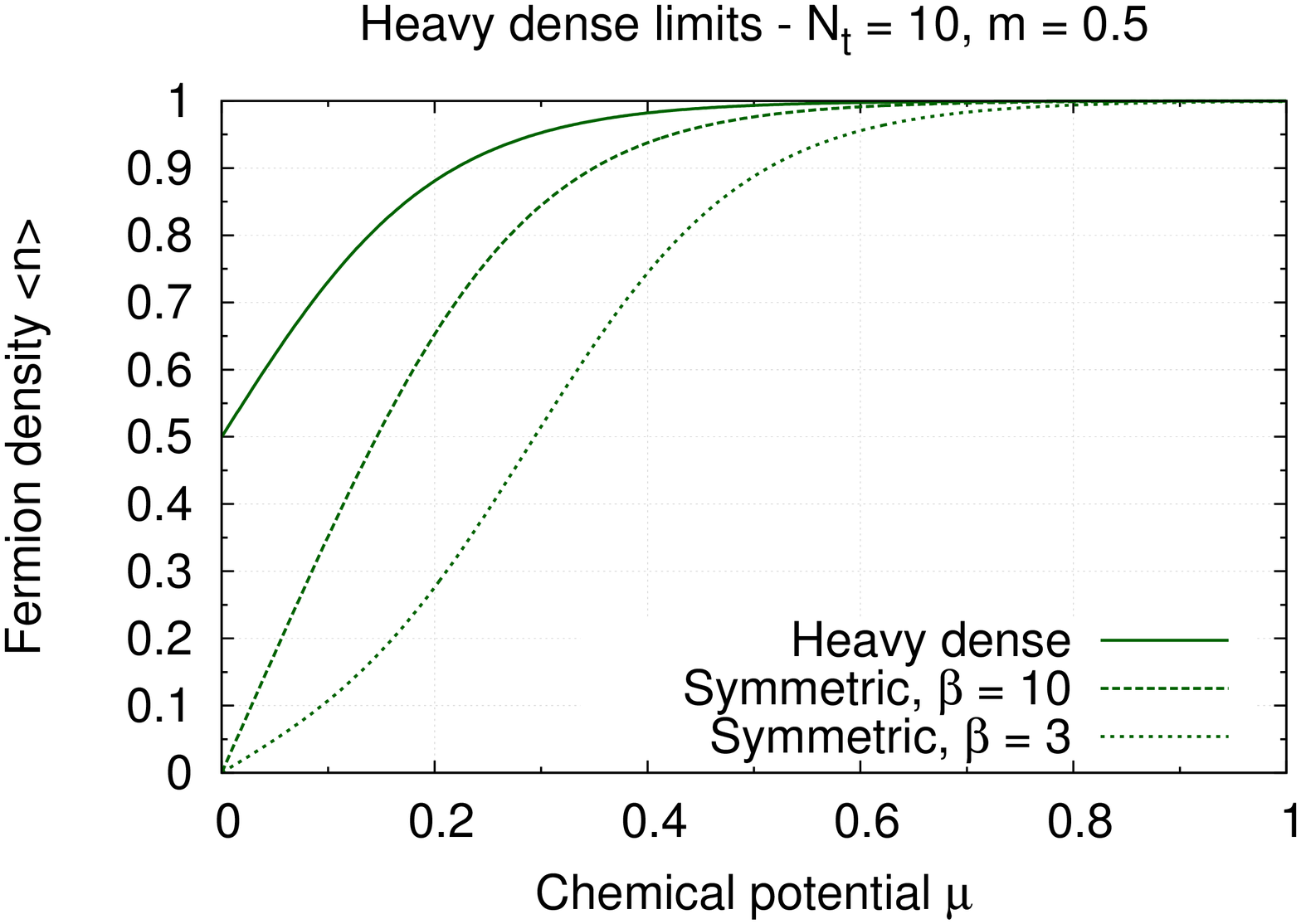}

\caption{\label{fig:Sym-HD} The density in the strict and in the symmetric
heavy dense limit.}
\end{figure}

The heavy dense limit given by \eqref{eq:Def-HD-limit} violates the
relation in \eqref{eq:Fermion-det-symmetry}. In the following, we
consider a model which restores the determinant symmetry but is not
a strict limit of the generalized Thirring model. Besides the Polyakov
loops in positive time direction with their characteristic $\exp\left(\mu\right)$
dependence, we will also consider the ones in negative Polyakov direction
with an $\exp\left(-\mu\right)$ dependence. In this sense we speak
of a symmetrized version of the heavy dense limit.

In this model the fermion determinant $\det\mathfrak{K}$ has the
form
\begin{multline}
\det\mathfrak{K}=\prod_{\mathbf{x}}\left(1+\xi_{f}\prod_{t=1}^{N_{t}}\left(1+\ii A_{0}\left(\mathbf{x},t\right)\right)\right)\\
\times\prod_{\mathbf{x}}\left(1+\xi_{b}\prod_{t=1}^{N_{t}}\left(1-\ii A_{0}\left(\mathbf{x},t\right)\right)\right),
\end{multline}
where we introduce $\xi_{f}\equiv\left(\kappa e^{\mu}\right)^{N_{t}}$
and $\xi_{b}\equiv\left(\kappa e^{-\mu}\right)^{N_{t}}$, compare
with \eqref{eq:Heavy-Dense-det}. The respective partition function
reads
\begin{align}
\mathcal{Z}_{1}^{\textrm{sym}} & =\intop_{-\infty}^{\infty}\prod_{x,v}\textrm{d}A_{v}\left(x\right)\,\det\mathfrak{K}\, e^{-S_{A}}\nonumber \\
 & =\frac{1}{\left(2\kappa\right)^{\Omega}}\left(\frac{2\pi}{\beta}\right)^{\frac{\Omega d}{2}}\left(1+\xi_{f}+\xi_{b}+\Delta_{\textrm{sym}}\right)^{V}
\end{align}
with shorthand notation
\begin{equation}
\Delta_{\textrm{sym}}=\kappa^{2N_{t}}\left(1+\frac{1}{\beta}\right)^{N_{t}}.
\end{equation}
For small $\mu$ the observables now receive additional contributions.
The density reads
\begin{equation}
\left\langle n\right\rangle =\frac{\xi_{b}-\xi_{f}}{1+\xi_{f}+\xi_{b}+\Delta_{\textrm{sym}}}
\end{equation}
and vanishes exactly at $\mu=0$. The fermion condensate is given
by
\begin{equation}
\left\langle \overline{\chi}\chi\right\rangle =\frac{2\kappa\left(1-\Delta_{\textrm{sym}}\right)}{1+\xi_{f}+\xi_{b}+\Delta_{\textrm{sym}}}
\end{equation}
and the equation of state takes the form
\begin{equation}
P=\log\left[\frac{\left(1+\xi_{f}+\xi_{b}+\Delta_{\textrm{sym}}\right)^{1/T}}{2\kappa}\left(\frac{2\pi}{\beta}\right)^{d/2}\right].
\end{equation}
The quantitative differences between the ordinary and the symmetric
heavy dense limits are small in the regime of large fermion masses
and large chemical potentials due to the exponential suppression of
the additional contributions. For small $m$ they alter the behavior
of observables massively, see Fig.~\ref{fig:Sym-HD}. However, for
very small $m$ the results again remain unphysical. Spatial contributions
to the fermion determinant become important, thus rendering the model
invalid.

\subsection{Analyticity in $\mu^{2}$}

For observables which are even in $\mu$ we can consider the analytic
continuation to $\mu^{2}<0$, corresponding to an imaginary chemical
potential. An example for such an observable is the fermion condensate
$\left\langle \overline{\chi}\chi\right\rangle $. Due to the relation
in \eqref{eq:Fermion-det-symmetry} the theory is free of a sign problem
in this case and we can make use of a real Langevin evolution. For
$\mu^{2}>0$ we employ a complex Langevin evolution.

Assuming that the complex Langevin evolution is correct, $\left\langle \overline{\chi}\chi\right\rangle $
should be analytic in $\mu^{2}$. Then any nonanalytical behavior
at $\mu^{2}=0$ has to be caused by incorrect convergence. We will
check this criterion in Sec.~\ref{sub:Consistency-conditions-numerical},
compare also with \cite{Aarts:2011zn}.

\subsection{Consistency conditions \label{sub:Consistency-conditions-theory}}

In \cite{Aarts:2011ax,Aarts:2011sf,Aarts:2009uq} the authors derived
an infinite tower of identities able to indicate the correctness of
complex Langevin evolutions. They could show that for all entire holomorphic
observables $\mathcal{O}$ the relation $\left\langle L_{\nu}\left(x\right)\mathcal{O}\right\rangle =0$
holds if and only if the evolution is converging correctly. Here
\begin{equation}
L_{\nu}\left(x\right)=\left(\frac{\textrm{d}}{\textrm{d}A_{\nu}\left(x\right)}+D_{\nu}\left(x\right)\right)\frac{\textrm{d}}{\textrm{d}A_{\nu}\left(x\right)}
\end{equation}
is the Langevin operator (summation over $\nu$ is not implied in
this subsection). A simple finite set of observables is defined by
$\mathcal{O}_{\nu}\left(x,k\right)=\exp\left(\ii kA_{\nu}\left(x\right)\right)$
with $k\in\mathbb{Z}$, yielding a necessary condition for correctness.
In Sec.~\ref{sub:Consistency-conditions-numerical} we will check
if
\begin{equation}
\left\langle L_{\nu}\mathcal{O}_{\nu}\right\rangle =\left\langle \ii k\left[\ii k+D_{\nu}\left(x\right)\right]e^{\ii kA_{\nu}\left(x\right)}\right\rangle =0
\end{equation}
holds for a wide range of parameters.

\section{Numerical Results \label{sec:Numerical-Results}}

\subsection{Implementation}

We solve the Langevin equation numerically using the adaptive stepsize
algorithm described in Sec.~\ref{sub:Formulating-the-Langevin-Eq}.
The implementation uses the GNU Scientific Library \cite{gough2009gnu}.
The code allows the calculation of the density, the condensate, the
phase factor of the fermion determinant, and the consistency conditions
of Sec.~\ref{sub:Consistency-conditions-numerical} for the full
generalized Thirring model. Furthermore, there is the option to carry
out calculations at purely imaginary chemical potential and in the
phase-quenched case.

To numerically determine observables we begin with a hot start, i.e.,
we randomly initialize the auxiliary field. This is followed by $5000$
thermalization steps. To evaluate the observables we sample the field
configuration typically $\mathcal{O}\left(10^{4}\right)$ times. Like
in the $\left(0+1\right)$--dimensional case \cite{Pawlowski:2013pje},
we estimate the error with a bootstrap analysis \cite{DeGrand:2006zz},
as the resulting error bounds generally prove to be more reliable.
However, in some cases we observe that the actual error is underestimated.

\subsection{Comparison to phase-quenched case}

\begin{figure}[p]
\subfloat[Density $\left\langle n\right\rangle $.]{\includegraphics[width=1\columnwidth]{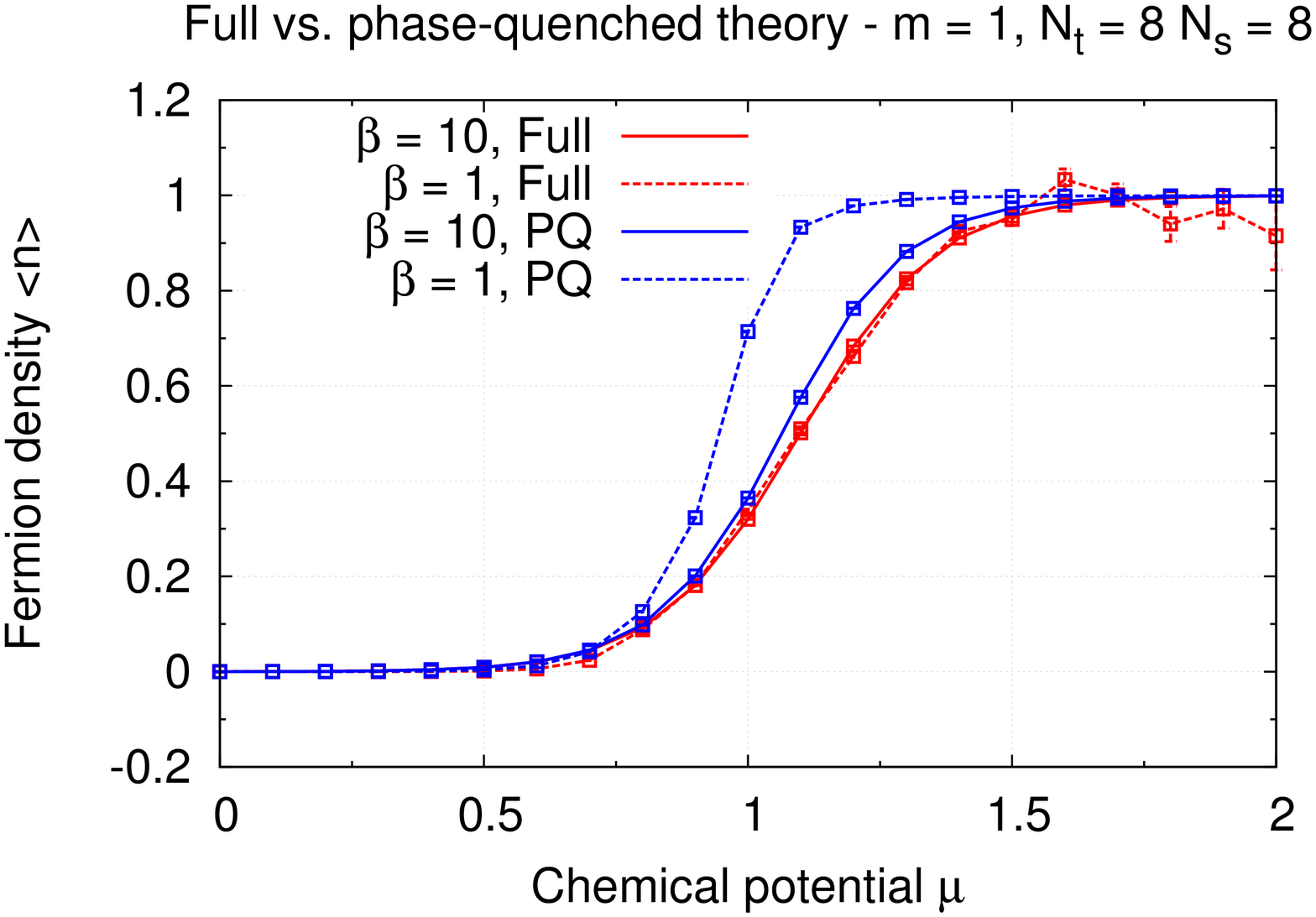}

}

\subfloat[Condensate $\left\langle \overline{\chi}\chi\right\rangle $.]{\includegraphics[width=1\columnwidth]{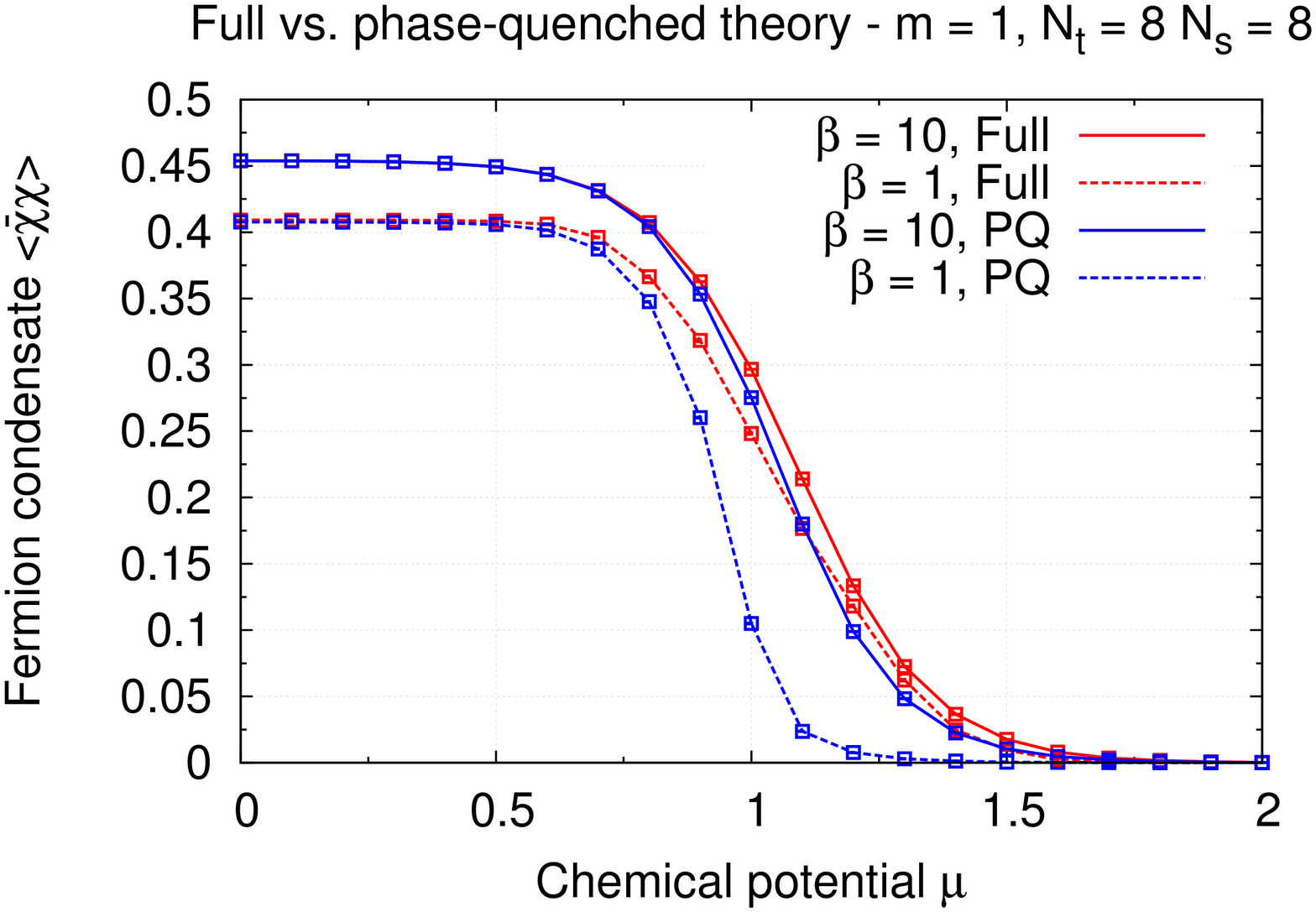}

}

\caption{Full and phase-quenched numerical results for $\beta=1,10$. \label{fig:Full-vs-pq}}
\end{figure}

In Fig.~\ref{fig:Full-vs-pq} we find a typical example for numerical
results of the density and the condensate in the full as well as the
phase-quenched case. The error bars are included, but are often too
small to be spotted with the naked eye. Although we only have $N_{t}=8$
time slices, we can already find hints for Silver Blaze behavior.
The onset to the condensed phase for the full theory generally lies
higher than in the phase-quenched case.

\begin{figure}
\includegraphics[width=1\columnwidth]{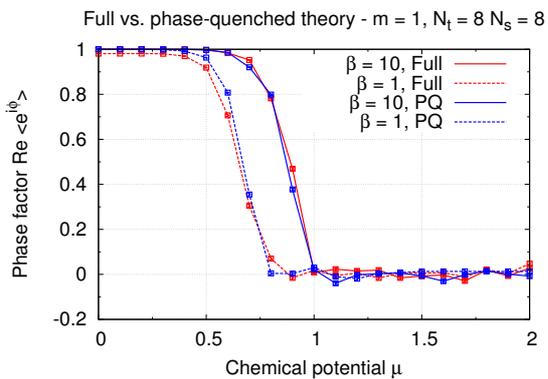}

\caption{Severity of the sign problem. \label{fig:Phase-factor}}
\end{figure}

The phase factor of the fermion determinant can serve as a measure
for the severity of the sign problem and can be found in Fig.~\ref{fig:Phase-factor}.
We see that for large $\beta$ it is less pronounced, but still quickly
approaches zero for increasing $\mu$.

\subsection{Evaluation in the heavy dense limit}

\begin{figure}
\subfloat[For mass $m=10^{2}$.]{\includegraphics[width=1\columnwidth]{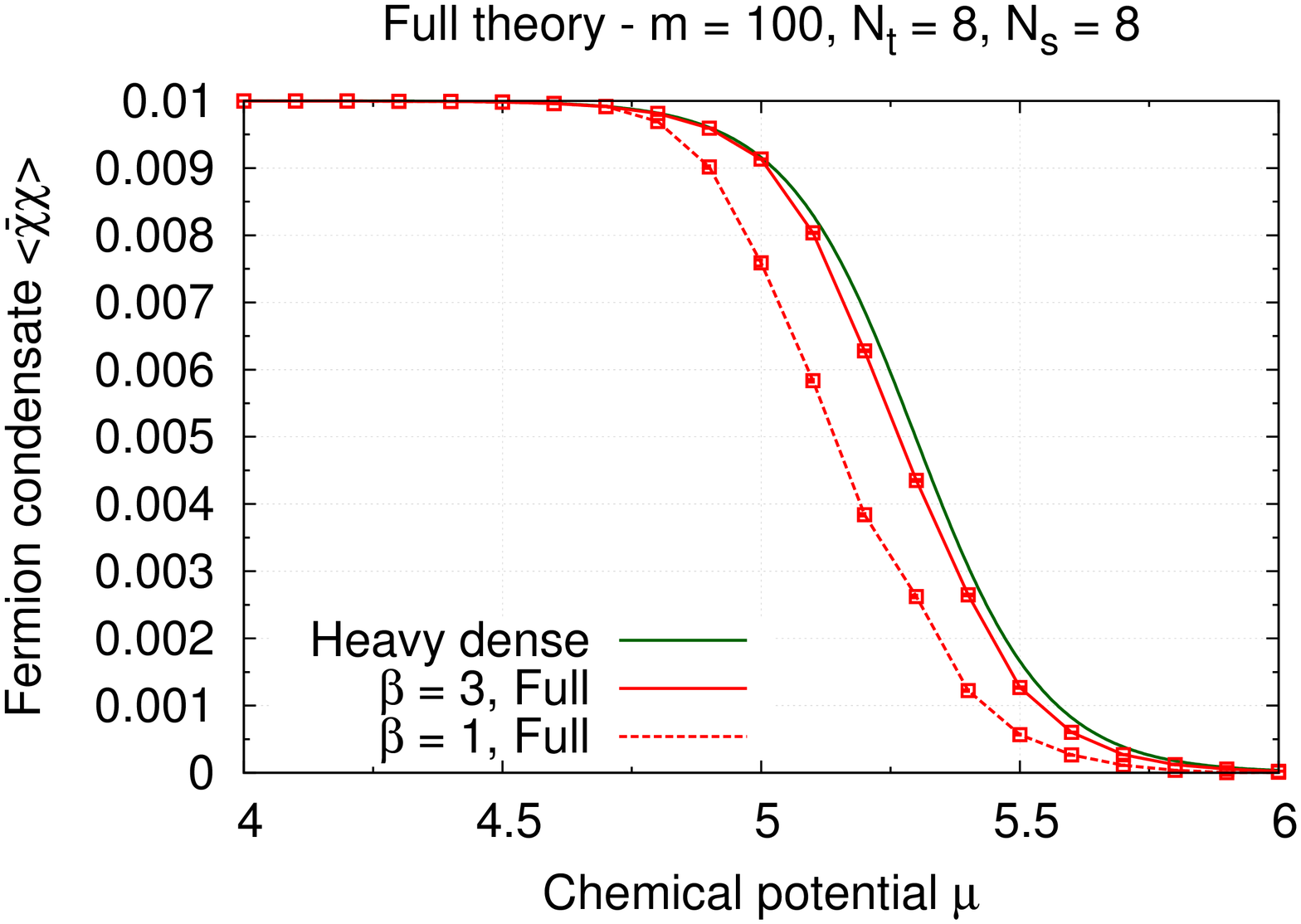}

}

\subfloat[For mass $m=10^{4}$.]{\includegraphics[width=1\columnwidth]{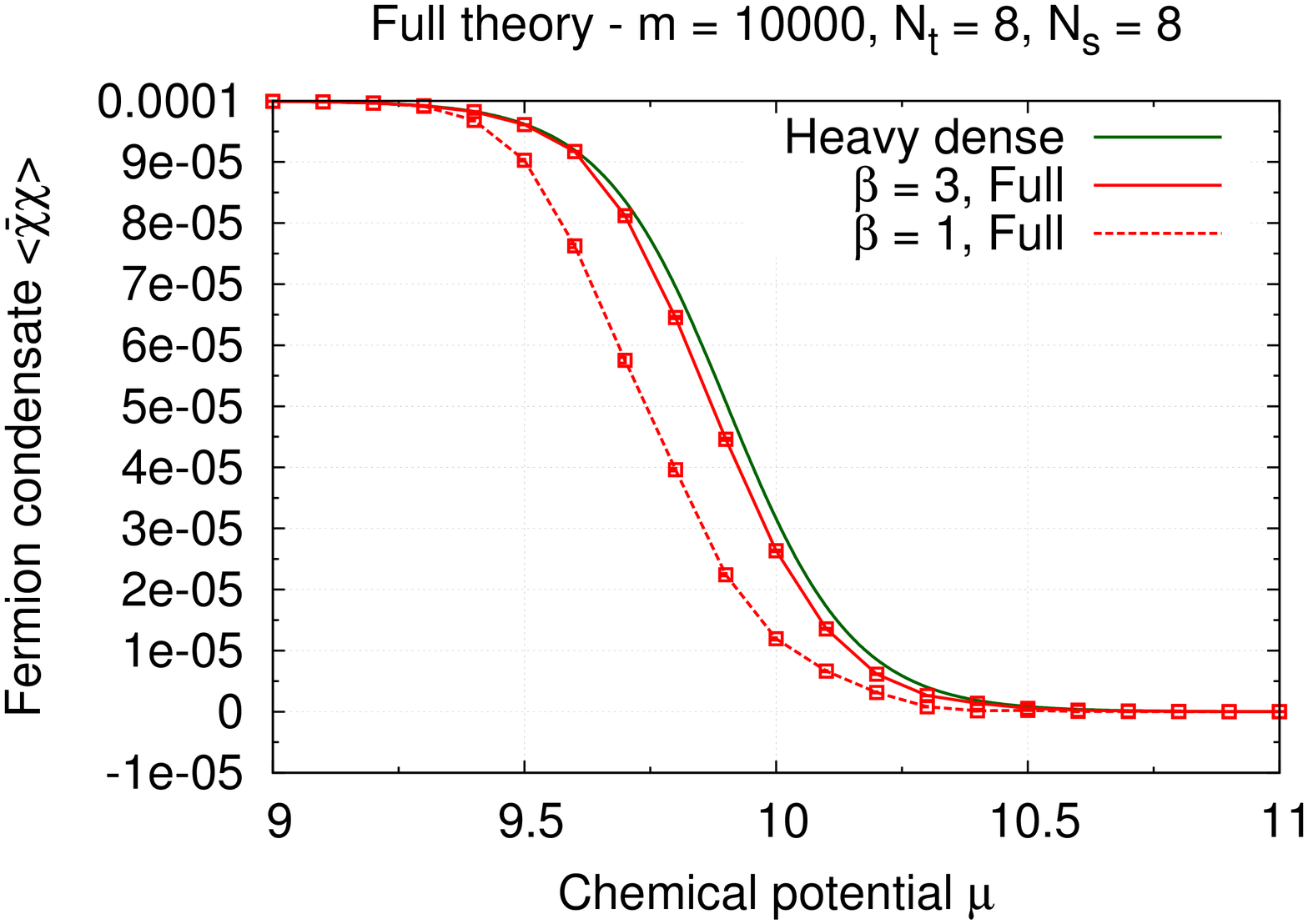}

}

\caption{Numerical results and heavy dense limit of the condensate for large
masses. \label{fig:Full-vs-HD}}
\end{figure}

In the analytical heavy dense limit in Sec.~\ref{sub:Single-flavor-partition-function},
the coupling factorizes, thus rendering the density and the condensate
independent of $\beta$. However, in Fig.~\ref{fig:Full-vs-HD} we
can see that the numerical results for the full theory show a dependence
on $\beta$ even for extremely large masses. Furthermore, the gap
between numerics and the analytical heavy dense limit  results seem
to persist even when increasing $m$ by several magnitudes. For smaller
$\beta$ this deviation is more pronounced, while for increasing $\beta$
we can see how the curves are approaching each other asymptotically,
similar to the $\left(0+1\right)$--dimensional case \cite{Pawlowski:2013pje}.

It is difficult to separate the different contributions to this gap,
as we are comparing an approximate analytical result with an algorithm
for the full generalized Thirring model, whose correctness we aim
to check. In the $\left(0+1\right)$--dimensional case we already
found a real deviation to theoretical results for small $\beta$,
and something comparable can potentially also happen in the $\left(2+1\right)$--dimensional
case.

\subsection{Coupling parameter dependence}

\begin{figure}
\subfloat[Fermion density $\left\langle n\right\rangle $.]{\includegraphics[width=1\columnwidth]{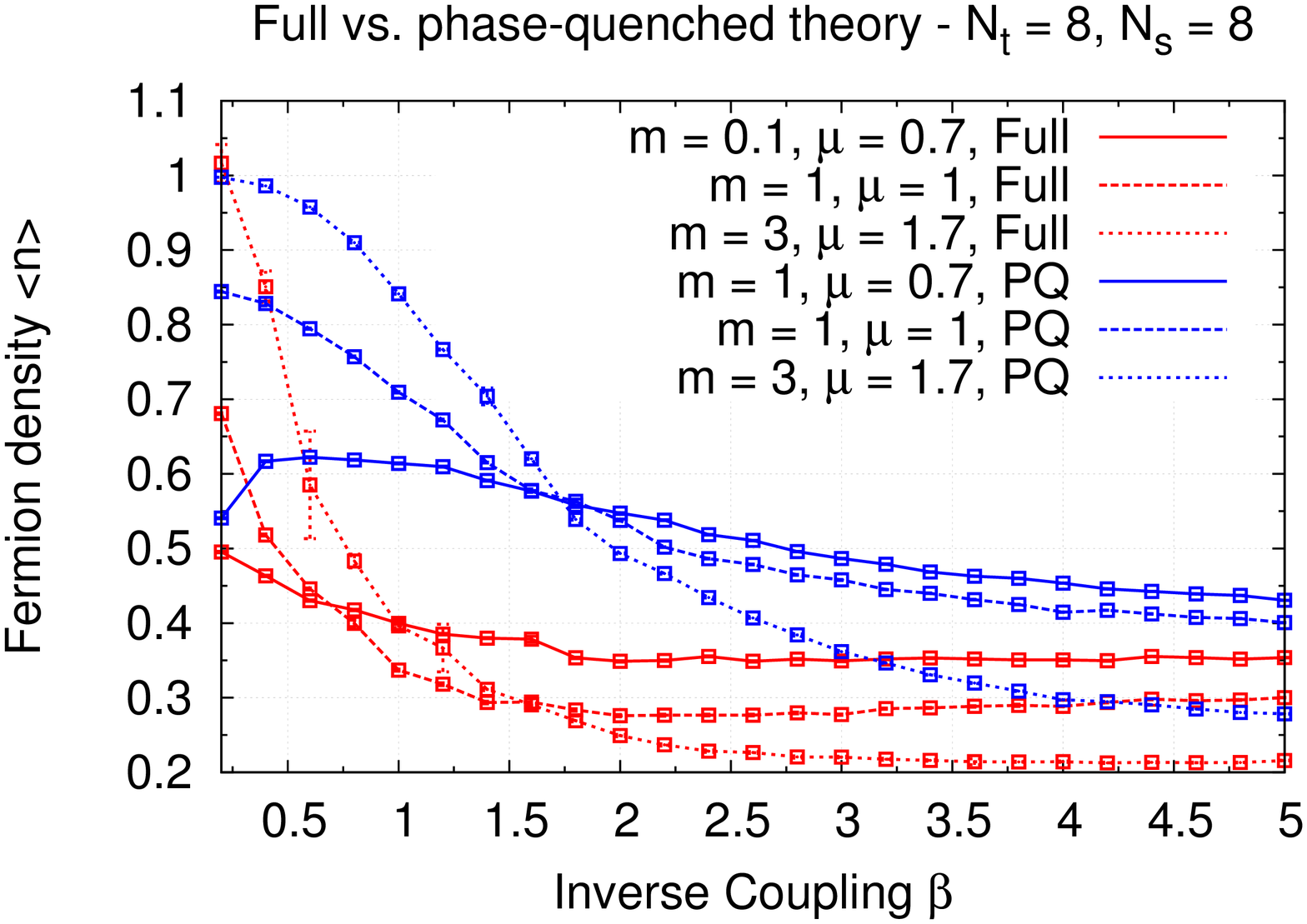}

}

\subfloat[Fermion condensate $\left\langle \overline{\chi}\chi\right\rangle $.]{\includegraphics[width=1\columnwidth]{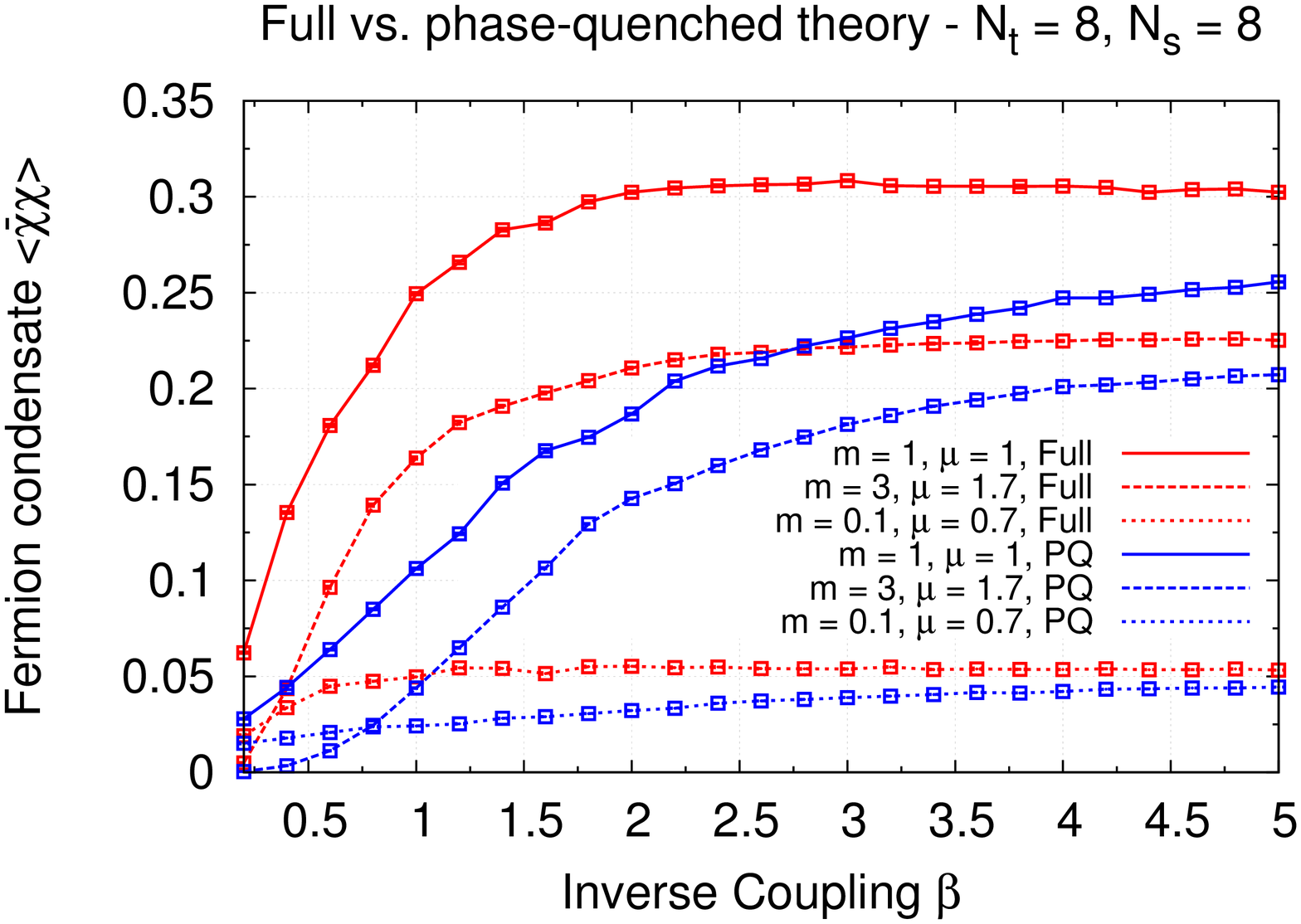}

}

\caption{\label{fig:Beta-Dependence} Observables as functions of $\beta$.}
\end{figure}

In Fig.~\ref{fig:Beta-Dependence} we plotted the numerically evaluated
observables as functions of $\beta$. For small $\beta$ it is more
difficult to obtain reliable results due to spikes caused by numerical
instabilities. For very large $\beta$ the results from the full theory
approach the phase-quenched calculations. As the heavy dense limit
makes no prediction about the $\beta$ dependence, the interpretation
of the findings is difficult.

\subsection{Several flavors \label{sub:Multi-flavor-numerical}}

\begin{figure}
\subfloat[Fermion density $\left\langle n\right\rangle $.]{\includegraphics[width=1\columnwidth]{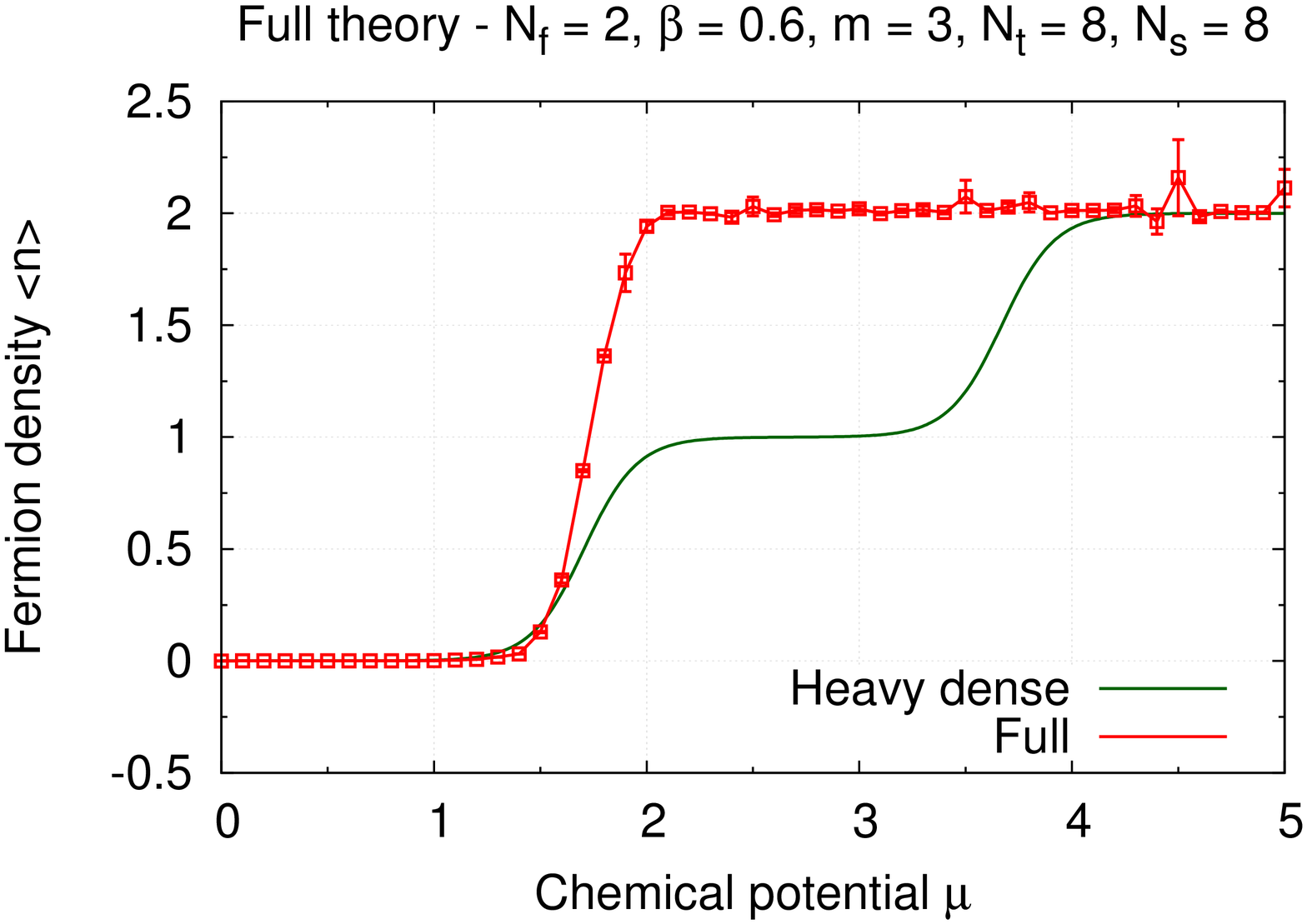}

}

\subfloat[Fermion condensate $\left\langle \overline{\chi}\chi\right\rangle $.]{\includegraphics[width=1\columnwidth]{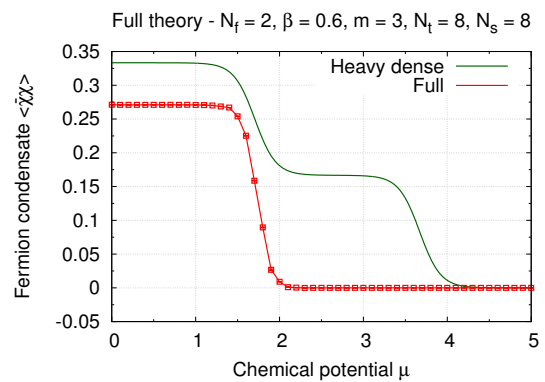}

}

\caption{\label{fig:Multiflavor} Observables for $\mathcal{N}=2$ degenerated
flavors.}
\end{figure}

In Fig.~\ref{fig:Multiflavor} we can find the density and the condensate
in the case of $\mathcal{N}=2$ staggered fermion flavors at a coupling
of $\beta=0.6$. As we saw in Secs.~\ref{sub:Partition-function-two-flavors}
and \ref{sub:Partition-function-many-flavors}, in a certain range
of $\beta$ these observables show up to $\mathcal{N}-1$ plateaus
in the analytical heavy dense results. However, we observe that the
numerical results for the full theory do not reproduce these plateau
structures at any value of $\beta$. Only when considering very large
$\beta$ is the theoretical curve free of these structures, and the
complex Langevin evolution approaches the correct result. This finding
is in agreement with \cite{Pawlowski:2013pje}.

\subsection{Consistency conditions \label{sub:Consistency-conditions-numerical}}

The numerical check of the consistency conditions of Sec.~\ref{sub:Consistency-conditions-theory}
for the $\mathcal{O}_{\nu}\left(x,k\right)$ observable yields several
violated conditions. Without loss of generality we restrict ourselves
to the case of $x=\left(1,1,1\right)^{\intercal}$. For improved statistics
we consider a $6^{3}$ lattice, where we sample the field configuration
$5\times10^{4}$ times. As an example we quote here
\begin{align}
\myRe\left\langle L_{0}\left(x\right)\mathcal{O}_{0}\left(x,k\right)\right\rangle  & =0.2258\pm0.0192,\\
\myRe\left\langle L_{1}\left(x\right)\mathcal{O}_{1}\left(x,k\right)\right\rangle  & =-0.0244\pm0.0049
\end{align}
for $\mathcal{I}=0$, $\mathcal{N}=\beta=m=\mu=k=1$ for a violated
condition. On the other hand,
\begin{align}
\myRe\left\langle L_{0}\left(x\right)\mathcal{O}_{0}\left(x,k\right)\right\rangle  & =0.0080\pm0.0058,\\
\myRe\left\langle L_{1}\left(x\right)\mathcal{O}_{1}\left(x,k\right)\right\rangle  & =-0.0004\pm0.0059
\end{align}
for $\mathcal{I}=0$, $\mathcal{N}=m=\mu=k=1$, $\beta=10$ is an
example for a condition which is compatible with a vanishing value.
As usual, $\mathcal{N}$ denotes the number of staggered fermion flavors,
$\mathcal{I}$ the imaginary noise, $\beta$ the inverse coupling
constant, $m$ the mass and $\mu$ the chemical potential.

If we take the error bounds estimated from the bootstrap analysis
seriously, we observe several violations of the consistency conditions
in $2+1$ dimensions. This is in agreement with the $\left(0+1\right)$--dimensional
case, where many consistency conditions seem to be unfulfilled. We
take this as a hint that the complex Langevin evolution in this canonical
implementation might not converge correctly.

\subsection{Imaginary noise}

\begin{figure}
\includegraphics[width=1\columnwidth]{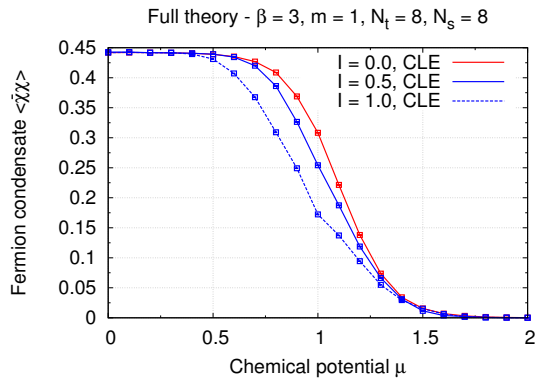}

\caption{Condensate $\left\langle \overline{\chi}\chi\right\rangle $ with
imaginary noise $\mathcal{I}$. \label{fig:Imag-noise}}
\end{figure}

We discussed in Sec.~\ref{sub:Formulating-the-Langevin-Eq} how we
can generalize from a real to an imaginary noise term, assuming correct
convergence of the complex Langevin evolution. In this case observables
should turn out to be independent of the noise term $\mathcal{I}$.
However, in Fig.~\ref{fig:Imag-noise} we can find clear evidence
for a dependence. Furthermore, we observe for almost all $\mathcal{I}>0$
a significantly larger deviation from analytical results. Hence, it
was also not possible to fine-tune $\mathcal{I}$ to a value so that
the complex Langevin evolution yields correct results.

\subsection{Analyticity in $\mu^{2}$}

\begin{figure}[t]
\subfloat[Mass $m=1$ at varying $\beta$.]{\includegraphics[width=1\columnwidth]{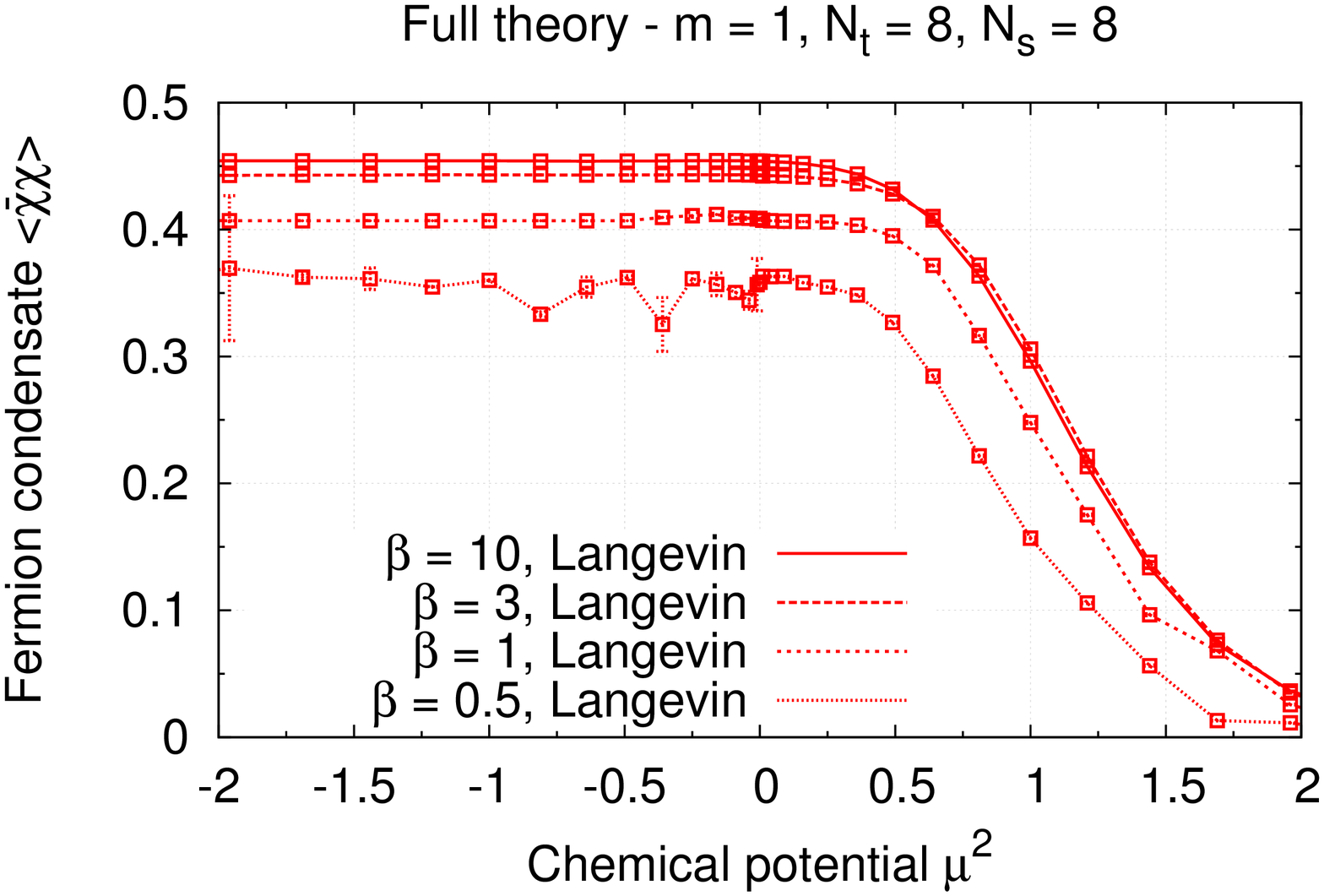}

}

\subfloat[Mass $m=0.2$ at $\beta=3$.]{\includegraphics[width=1\columnwidth]{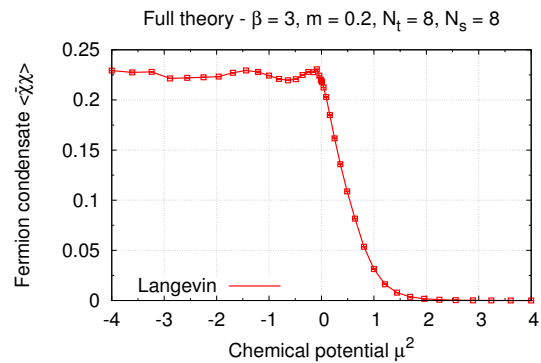}

}

\caption{Condensate $\left\langle \overline{\chi}\chi\right\rangle $ as a
function of $\mu^{2}$. \label{fig:Analyticity-mu2}}
\end{figure}

In Fig.~\ref{fig:Analyticity-mu2} we find the condensate $\left\langle \overline{\chi}\chi\right\rangle $
as a function of $\mu^{2}$. While for $\mu^{2}>0$ the numerical
evaluation employs a complex Langevin evolution, for $\mu^{2}\leq0$
we use a real one. For small $\beta$ the numerical evaluation tends
to be unstable. Like in the $\left(0+1\right)$--dimensional case
$\left\langle \overline{\chi}\chi\right\rangle $ is analytic within
the numerical accuracy. This suggests that the method can work sufficiently
well for small $\mu$.

\section{Conclusions \label{sec:Conclusions}}

In this paper we have extended our $\left(0+1\right)$--dimensional
investigation of the generalized Thirring model in \cite{Pawlowski:2013pje}
to the $\left(2+1\right)$--dimensional case. The numerical findings
are similar to those obtained in $0+1$ dimensions. The complex Langevin
evolution allows us to evaluate observables in the full theory in
a straightforward way, despite the severe sign problem. However, our
investigation suggests that in some cases it does not converge towards
the physical theory. This was indicated by several violated consistency
conditions, a gap in analytical predictions in the heavy dense limit
and the absence of plateaus in the observables for $\mathcal{N}>1$
flavors.

Despite these observations, we found the condensate $\left\langle \overline{\chi}\chi\right\rangle $
to be analytical at $\mu^{2}=0$, suggesting that a complex Langevin
evolution seems to work for small $\mu$. Also, for large $\beta$
the heavy dense limit results could be reproduced. In these regimes
we then have an appealing method to tackle the sign problem in the
generalized Thirring model.

Further investigations have to deal with the question of how to address
the aforementioned problems. In particular, coordinate transformations
as suggested in \cite{Aarts:2012ft} and gauge-cooling procedures
like the one employed in \cite{Seiler:2012wz} might allow a stabilization
of the complex Langevin evolution.

\section*{Acknowledgments}

We thank I.-O.~Stamatescu for many helpful remarks and clarifying
discussions. We thank G.~Aarts, D.~Sexty and E.~Seiler for discussions.
This work is supported by the Helmholtz Alliance Grant No.~HA216/EMMI
and by Grand No.~ERC-AdG-290623. C.~Z.~thanks the German National
Academic Foundation for financial support.

\bibliography{literature}

\end{document}